\documentclass[9pt,twocolumn,twoside]{preprint}
\templatetype{pnasresearcharticle} 

\usepackage{graphicx}      %
\usepackage{microtype}
\usepackage{csquotes}

\title{MYND: Unsupervised Evaluation of Novel BCI Control Strategies on Consumer Hardware}

\author[a, 1]{Matthias R. Hohmann}
\author[a]{Lisa Konieczny}
\author[a]{Michelle Hackl}
\author[a]{Brian Wirth}
\author[b]{Talha Zaman}
\author[b]{\\Raffi Enficiaud}
\author[c]{Moritz Grosse-Wentrup}
\author[a]{Bernhard Schölkopf}
\affil[a]{Department for Empirical Inference, Max Planck Institute for Intelligent Systems, Max-Planck-Ring 4, 72076 Tübingen, Germany\\}
\affil[b]{Software Workshop, Max Planck Institute for Intelligent Systems, Max-Planck-Ring 4, 72076 Tübingen, Germany\\}
\affil[c]{Research Group Neuroinformatics, University of Vienna, Hörlgasse 6, A-1090 Vienna, Austria\\}
\affil[ ]{\{mhohmann, lkonienczny, mhackl, bwirth, tzaman, renficiaud, bs\}@tue.mpg.de, moritz.grosse-wentrup@univie.ac.at}

\leadauthor{Hohmann} 

\significancestatement{Using thoughts to interact with our environment could be the next frontier of medical and personal technology. The deployment of brain-computer interfaces (BCIs) on affordable hardware in daily life is the ultimate goal of this field, and both researchers and users need novel ways to evaluate conceptual systems under realistic conditions. We present a framework for the unsupervised evaluation of BCI control strategies on a consumer-grade electroencephalogram. We show that combining real-time guidance for hardware administration with a set of control strategies that induce broad spatial modulations in neural activity may be a viable basis for research on accessible BCI usage in daily life.}

\authorcontributions{M. R. H., M. G-W., and B. S. had the idea of building a platform for large-scale EEG recordings. 
M. R. H. designed and developed the platform, conducted experiments, analysed and interpreted the data, performed a literature review, and wrote the paper. 
M. H. and B. W. assisted with the design of the platform.
L. K., M. L. and B. W. assisted with experiments.
T. Z. and R. E. assisted with software development.
M. G-W. and B. S. provided advice on the platform development and supervised the work.
All authors commented on the manuscript.}
\authordeclaration{The authors declare no conflict of interest.}
\correspondingauthor{\textsuperscript{1}To whom correspondence should be addressed. E-mail: mhohmann@tue.mpg.de}


\begin{abstract}
Neurophysiological studies are typically conducted in laboratories with limited ecological validity, scalability, and generalizability of findings. 
This is a significant challenge for the development of brain-computer interfaces (BCIs), which ultimately need to function in unsupervised settings on consumer-grade hardware.
We introduce \textit{MYND}: A framework that couples consumer-grade recording hardware with an easy-to-use application for the unsupervised evaluation of BCI control strategies. 
Subjects are guided through experiment selection, hardware fitting, recording, and data upload in order to self-administer multi-day studies that include neurophysiological recordings and questionnaires. 
As a use case, we evaluate two BCI control strategies (\enquote{Positive memories} and \enquote{Music imagery}) in a realistic scenario by combining \textit{MYND} with a four-channel electroencephalogram (EEG).
Thirty subjects recorded 70.4 hours of EEG data with the system at home. 
The median headset fitting time was 25.9 seconds, and a median signal quality of 90.2\% was retained during recordings.
Neural activity in both control strategies could be decoded with an average offline accuracy of 68.5\% and 64.0\% across all days. 
The repeated unsupervised execution of the same strategy affected performance, which could be tackled by implementing feedback to let subjects switch between strategies or devise new strategies with the platform.
\end{abstract}

\dates{This manuscript was compiled on \today}
\bibliography{/Users/mhohmann/Documents/library.bib}
\begin{document}

\maketitle
\thispagestyle{firststyle}
\ifthenelse{\boolean{shortarticle}}{\ifthenelse{\boolean{singlecolumn}}{\abscontentformatted}{\abscontent}}{}

\begin{figure*}[t]
  \centering
  \includegraphics[width=11.4cm]{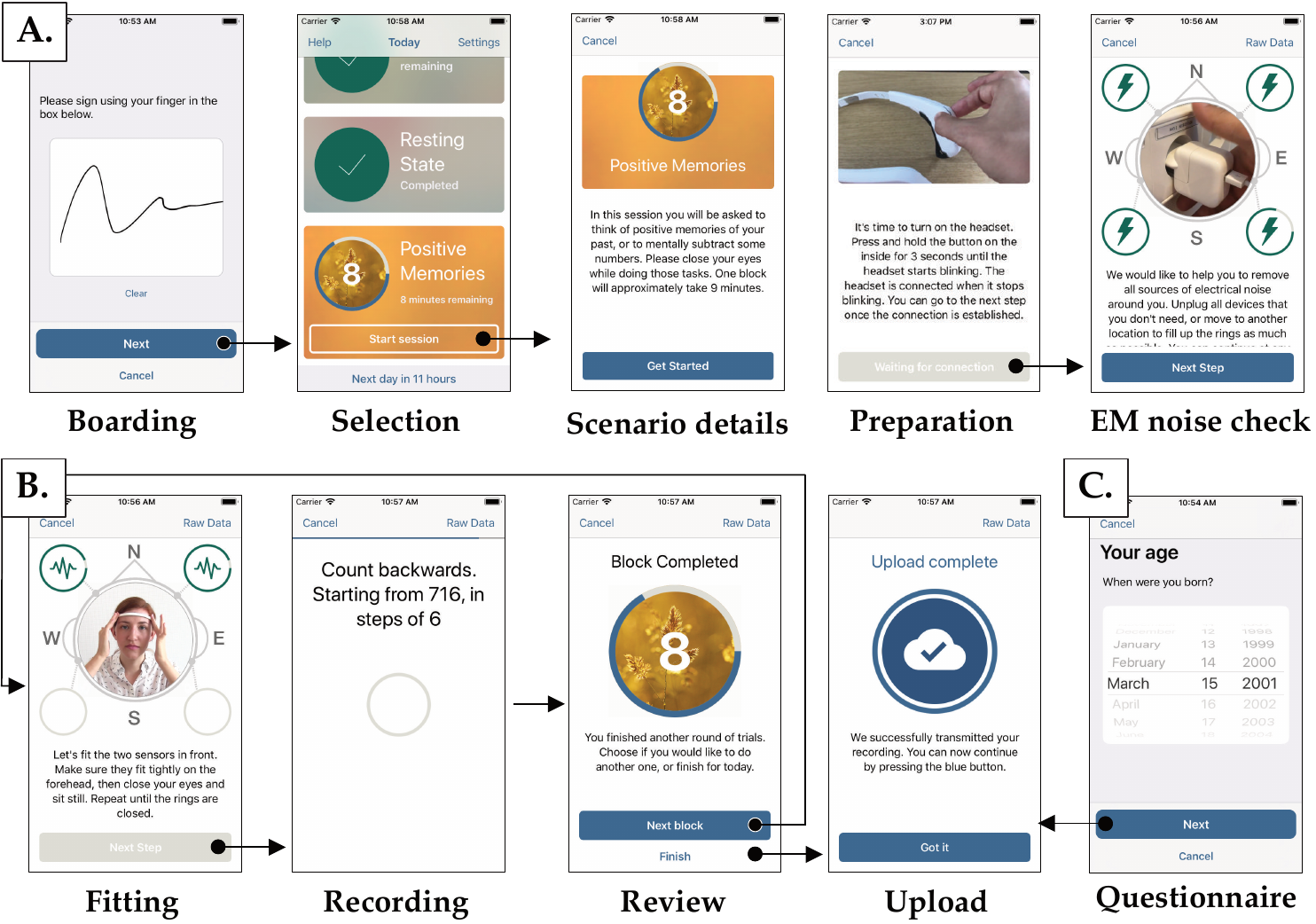}
  \caption{The user-interface flow of participating in a study with \textit{MYND}. 
  (A) After boarding and consent review, subjects can select a scenario, inform themselves about the scenario's tasks, prepare the hardware, and fit a suitable location with the electromagnetic (EM) noise check.
  (B) A scenario is broken up into several blocks of trials. 
  After each block, subjects can review their progress and choose to proceed to the next block with an intermediate assessment of the hardware's fit, or end the recording session and upload the recorded data.
  (C) An exemplary screen of a questionnaire scenario. Multiple-choice, short text answers, and date selection are supported.}
  \label{fig:workflow}
\end{figure*}

\dropcap{N}europhysiological research is typically bound to laboratory environments. Investigating the basics of neural communication requires complex and expensive setups, shielded as much as possible from environmental influences. 
It also requires the presence of an expert to supervise the equipment and the experimental procedure. 
While neurophysiological studies have advanced our understanding of the human brain over the past decades, the controlled and expensive nature of laboratory research makes this track of research less suitable to capture contextual influences that would only occur in daily life \citep{Spooner2006}. 
The validation of findings in a broader, more diverse population is often logistically infeasible \citep{Button2013}. 
This limitation is particularly profound in research on brain-computer interfaces (BCIs), which aims to translate cortical signals into computer commands for everyday communication, control, or treatment. 
Many BCIs rely on a control strategy: A set two or more tasks that induce differences in neural activity. 
Those tasks could be executed as a proxy to say, for example,~\enquote{yes} or \enquote{no} with a BCI system. 
Controlled and expensive research with laboratory equipment may allow for investigating the induced cortical effects in detail, but it is less suitable for investigating self-supervised, long-term usage. 
The deployment of BCIs on affordable hardware in unsupervised scenarios is the ultimate goal of this field, and both researchers and users need novel ways to evaluate conceptual systems under realistic conditions.

This paper introduces \textit{MYND}: A framework that couples a consumer-grade electroencephalogram (EEG) with an easy-to-use application for the unsupervised evaluation of BCI control strategies.
For researchers, \textit{MYND} provides the option to add ecological validity and scalability to traditional lab-based BCI studies.
For participants, \textit{MYND} enables at-home participation in BCI studies without the assistance of an expert.
The application combines an easy-to-use interface with a real-time feedback algorithm to guide the fitting of a consumer-grade EEG. 
It allows for an immediate transfer of both neurophysiological data and questionnaire data to the researcher.
In multi-day studies, it enables subjects to choose recording times and the number of sessions to record on a given day within certain experimental constraints.

\subsection*{The current study}
We developed \textit{MYND} to evaluate BCI control strategies in a realistic scenario on consumer-grade hardware. 
In the current study, we present results for the two strategies \enquote{Positive memories} \citep{Hohmann:vc} and \enquote{Music imagery} \citep{GroHohPetGro17}.
Single-session laboratory studies showed that both strategies feature two tasks that modulate neural oscillations in broad areas of the parietal and prefrontal cortex. 
These broad spatial effects could make them particularly suitable for use on a consumer-grade, headband-like EEG with few sensors. 
Previously, we piloted a first version of \textit{MYND} at our institute \citep{Hohmann2019}.
We could show that eighteen subjects were able to use the application without direct supervision. 
Subjects recorded neurophysiological activity with visible differences between both tasks of the \enquote{Positive memories} strategy.
Now, we report the results of the first at-home study with 30 subjects that used \textit{MYND} over seven days. 
We make the following contributions:
\begin{itemize}
    \item We introduce MYND, a platform for unsupervised evaluation of BCI control strategies on consumer hardware.
    \item In terms of usability, we show that subjects self-administered the EEG headset with the implemented fitting procedure over several days, and they retained a high signal quality during tasks. %
    \item Regarding the unsupervised, daily use of both control strategies on consumer-grade hardware, we show that \enquote{Positive memories} and \enquote{Music imagery} induce differentiable neural activity across days.
    \item Concerning their employment in future BCIs, we found that repeated execution of the same strategy affected decoding accuracy. We discuss how feedback could enable subjects to switch between control strategies or devise new strategies in future research with the platform.
\end{itemize}

\subsection*{Related work}
BCI development is an area of research that is concerned with utilizing self-induced modulations in neural activity for communication, control, or treatment \citep{Wolpaw2002}.
Until recently, work in this field was mainly driven by clinical applications: 
The most prominent example is the prospect of developing a communication system for people that have complete paralysis through amyotrophic lateral sclerosis, which has yet to be realized reliably \citep{Marchetti2015, Spuler2019}.
Non-invasive BCIs translate the user's cortical signals, typically recorded with EEG hardware, into control signals through machine learning, and forward them to an assistive application that can be used for communication or control of external appliances.
Users are instructed to modulate neural activity such that it can be differentiated and mapped onto digital commands. 
We evaluate \enquote{asynchronous} control strategies with \textit{MYND}. 
Here, users are instructed to produce specific thoughts without reacting to external stimuli. 
BCI strategies that have been explored in this context include motor imagery \citep{Pfurtscheller2001}, spatial navigation, and mental calculation \citep{Friedrich2012}, music imagery \citep{GroHohPetGro17}, and daydreaming \citep{Hohmann2016}. 

Because of its roots in clinical neuroscience and engineering, methods and perspectives on BCI research are still based on traditional research paradigms.
A combination of BCIs and human-computer interaction research has mostly been done to augment existing interaction paradigms.
In passive BCIs, an existing, conventional interaction paradigm is augmented with background EEG recordings in order to adapt them to workload or emotional correlates \citep{Zander2016,Arico2018}.
For clinical purposes, \citep{Murphy2018} combined background recordings of a custom EEG prototype with a smartphone-based memory and reaction tasks to detect early signs of cognitive impairment due to dementia. 
However, actively using BCIs for communication and control remains a significant challenge in the field.
Hardware and software for end-users constitute a notable barrier in BCI development \citep{Huggins2019}.
BCIs often employ expensive biomedical research equipment as recording hardware and software, which can rarely be afforded or used by consumers and requires expert knowledge and laboratory environments.
Several attempts have been made to tackle this limitation:
OpenBCI\footnote{https://www.openbci.com} offers Bluetooth-enabled amplifiers, electrodes, and customizable headwear to create low-cost EEG systems. 
Laboratory studies have employed this technology (e.g.,~\cite{Xu2018}), and it could be used as a basis for open and easy-to-use technology for at-home studies and BCI development.
Others have proposed the merging of consumer hardware and research hardware \citep{DeVos2014}, the development of smaller electrodes that can be attached behind the ear \citep{Debener2015}, or in-ear solutions \citep{Goverdovsky2016}. 
However, these attempts are still in early stages of development, and, at the time of this study, lacked ergonomic properties and software support that would make them suitable for unsupervised at-home recordings across several sessions.

Apart from more developer-oriented approaches, several \enquote{direct-to-consumer} EEGs emerged recently. 
These systems are typically accompanied by smartphone software that is designed for meditation assistance and self-quantification \citep{Ienca2018}.
The Muse EEG\footnote{InteraXon, Canada, https://choosemuse.com}, which is used for the current study, provides a smartphone application that assists users with guided meditation and feedback on attention and relaxation levels.
The Dreem EEG\footnote{Dreem.co, France, https://dreem.com/}, a headset that is specifically designed to be worn at night, offers an application that analyses and classifies sleep patterns through an online service. 
Headsets by Emotiv\footnote{Emotiv, USA, https://www.emotiv.com} feature between five and fourteen channels and are bundled with software subscriptions for private or business use.
These devices are designed for self-administration and longer recording sessions.
Through custom software, several of them have been used in laboratory studies: 
In \citep{Krigolson2017} the Muse EEG was successfully used for a spelling system, noting its ease-of-use and sufficient data quality for a BCI application.
Emotiv headsets have been used in several BCI studies as well \citep{Duvinage2013,Liu2012}. 
The Dreem EEG and its bundled proprietary sleep classification platform are promoted as a research tool for sleep studies by the developer \citep{Arnal662734}.
However, the bundled end-user applications are not intended for BCI research and do not expose processing or data storage interfaces.
The custom tools that are developed by researchers to interface with these consumer-grade systems are typically not meant for use by non-experts outside of the laboratory.

The gap between consumer- and research-technology motivated the development of \textit{MYND}: 
Several promising tools for BCI evaluation in realistic scenarios exist, but they are in early stages of development and geared towards developers and researchers. 
On the other hand, consumer-grade EEG systems are optimized for usability, but they are bundled with smartphone applications that are specifically designed for private use and self-quantification. 
With \textit{MYND}, we aim to complement laboratory BCI research with a smartphone application that allows for questions about subjective experience, different environments, and longitudinal use, which can hardly be captured with traditional paradigms.

\section*{System design and implementation}
We devised our initial requirements for \textit{MYND} based on our experience with laboratory-based BCI research: 
The platform needs to assist with the self-administered fitting of the recording hardware. 
Consent and questionnaire forms should be digitally implemented on the device. 
All recorded data should be transmitted in an encrypted format. However, study progression and processing should never be dependent on internet access in order to guarantee participation from wherever subjects are located.
We also implemented multi-language and multi-day support, as well as an electromagnetic noise detection to assist subjects with finding favorable locations for recordings.

Participating in a neurophysiological study on \textit{MYND} features six steps: (1) the initial boarding and consent review, (2) the selection of an experimental scenario, (3) hardware preparation, (4) hardware fitting, (5) recording data, and, when a scenario was completed, (6) storage and upload of the recorded data.  
In this section, we will describe the implementational details of each step. \textit{MYND} is written in Swift 4.2 for iOS.
Figure \ref{fig:workflow} illustrates the following steps.

\subsection*{Boarding a study}
First, subjects board a study by reviewing the consent form. We utilized ResearchKit\footnote{Apple Inc., USA, http://researchkit.org} to display the various consent sections and obtain a hand-written signature from participants. The consent is transmitted as a PDF.   
For accessibility, subjects can choose to enable text-to-speech (TTS) for all instructions and questionnaires. 
TTS ensures that subject groups with impaired vision can operate the application, and it allows subjects to perform tasks in a comfortable position without looking at the screen. 
Lastly, they can set up Internet connectivity and enable reminders for when to start the next day of the study.
Once the subject consented and adjusted their preferences, the application transitions to the home screen.

\subsection*{Selection of an experimental scenario}
An experimental scenario can either contain a questionnaire or a neurophysiological recording. 
Questionnaire and neurophysiological recording scenarios that should be completed on the current day of the study are shown. 
Numbers represent the remaining minutes of scenarios, and circles represent progress. 
Inactive scenarios are greyed-out, to give subjects a quick overview over completed and remaining scenarios without distracting them from the current scenarios.
Only the current scenario displays the \enquote{Start Session} button, which initiates the recording procedure. 
For announcements and technical support, text messages can be downloaded from a server in the background and displayed before the recording commences.

\subsubsection*{User-initiated time-outs}
Different from lab-based studies, recording times may be scattered during a day. 
There may also be days where no recording can be performed due to other obligations. 
In order to maintain control over recording times while still allowing for the flexibility to integrate recording sessions into daily life, we implemented a user-initiated time-out. 
A timer starts with the first recording that is performed on a day. Subjects are asked to perform all recordings for a given day within a twelve-hour timeframe. 
If a subject finishes all recordings, the application is locked until the timer expires. 
When the timer expires, the application loads the next set of sessions and sends an optional notification.

In the beginning, every scenario features a short description, an image, and information about the duration of the remaining blocks to be completed. 
The duration is estimated based on the number of questions or neurophysiological tasks in a scenario and a fixed estimate of the required preparation time.

\subsection*{Preparation}
Neurophysiological recording sessions begin with the initial preparation of the hardware. 
A short video complements the instruction in each step on top of the screen. 
Subjects are first asked to prepare their head for the headset by tying back hair and removing glasses.
Subjects are also asked to sprinkle water onto the parts of the head that the headset will rest on, as we discovered during pilot testing that this improves sensor conductivity.
Then, subjects are asked to turn on the headset, pull out the side arms, place it loosely on their head, and push the side arms back in to make it fit tightly. 

In general, all steps are self-paced. Subjects proceed to the next step by pressing the single blue button on the bottom of the screen. 
However, several steps include a mandatory condition that needs to be fulfilled in order to proceed. 
In these steps, the button is greyed out until the condition is met. 
The \enquote{turn on the device} step uses this feature to ensure that the EEG headset was found, and a Bluetooth connection was established before the subject proceeds. 
It is also used later in the fitting procedure to ensure that subjects met the required signal quality threshold before the recording starts.
When the condition is met, short acoustic feedback is given, and the button turns blue to indicate that the subject can proceed.

As an additional measure to prevent data loss, the application monitors the battery of the EEG headset.
Subjects can only proceed when more than 10\% of the headset battery is remaining. 
If the headset was not found, the battery of the headset is too low, or the headset disconnected due to other reasons, the current block is aborted with a message that describes the issue. 
The current block is also aborted, and the headset is disconnected when the application enters the background. This constraint was implemented to ensure that subjects look at a controlled view when recording data, and to prevent potential battery saving processes from disconnecting the headset or from impairing processing.

\subsubsection*{Environmental quality}
In the pilot study in \citep{Hohmann2019}, subjects sometimes placed themselves in the proximity of a power outlet or other appliances which induced electromagnetic (EM) noise that impaired the fitting procedure.
To assist subjects with finding a location that is least exposed to EM noise, we implemented a noise detection step.
Progress rings around a stylized head represent the respective electrodes of the recording hardware. Rings fill up with less EM noise. The application uses real-time signal processing to map the 50~Hz component of the incoming EEG time-series onto an environmental quality estimate between 0\% and 100\%. 
50~Hz is the AC frequency of electronic devices in Europe. 
Based on a visual inspection of recordings in different proximity to EM noise inducing appliances, we set a log-band-power of $-1\mu V^{2}$ to equal 100\% environmental quality.
The noise detection step is non-blocking, as subjects may not find a location at home that is sufficiently free of EM noise.

\subsection*{Fitting}
After a sufficiently noise-free environment was found, the application assists users with fitting the recording hardware properly. 
Instructions are first given on the frontal sensors, as those typically already achieve a good fit with the general adjustments of the headset.
Then instructions are given on the rear sensors. 
Similar to the noise-detection screen, progress rings represent the signal quality at each sensor.
Real-time signal processing maps the incoming EEG time-series onto a signal quality estimate between 0\% and 100\%. 
Subjects are required to reach 100\% signal quality before the recording commences. 
During pilot testing, we noticed that subjects would become frustrated after trying to fit the headset for more than three minutes. 
We drop the target signal quality to 75\% after three minutes of hardware fitting to keep subjects motivated and to allow them to proceed if a perfect fit cannot be achieved due to environmental or physiological constraints. 
We did not implement a maximum fitting time, so we could evaluate how long subjects would try to fit the headset without constraints.

The fitting algorithm uses the variance of the EEG signal to estimate its quality. As electromagnetic noise and movement artifacts are likely to induce more variation in the signal than cortical signals, a smaller total variance indicates fewer artifactual influences. This method has been proposed before in a laboratory setting \citep{Krigolson2017}, where experts inspected the raw EEG data and the variance computed over one-second time-windows to determine when signal quality was sufficient.
A variance threshold of $150\mu V^{2}$ or less was reported to reflect a sufficient signal quality.
However, apart from headset fit, the EEG signal is strongly affected by eye-blinks and other involuntary movements. While experts can ignore these artifacts when investigating the raw EEG signal, the variance computation is strongly affected. 
During early development, we found that merely relating the variance to a threshold would result in noisy feedback, as involuntary movements could suddenly increase variation. Additionally, a time-window of one second was too long, and the delay between adjusting the headset and an update of signal quality was frustrating to subjects. 

\begin{figure}[H]
  \centering
  \includegraphics[width=\columnwidth]{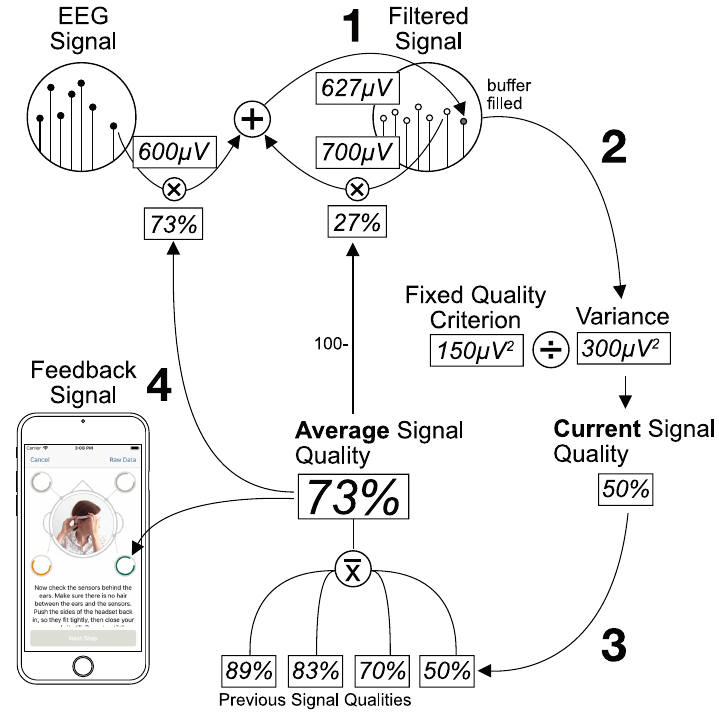}
  \caption{The signal processing pipeline for hardware fitting. 1. An incoming data point of the raw EEG signal is filtered and appended to a buffer. Filtering occurs by a weighted average with the previous data point of the filtered signal. 2. When the buffer of the filtered signal is filled, the variance is computed and compared to a fixed quality criterion. 3. The resulting signal quality estimate is then again buffered. 4. The average is used both for weighting the next incoming data points and the user-interface. Italic values in boxes represent example values.}~\label{fig:signalProcessing}
\end{figure}

Therefore, the algorithm filters the raw EEG signal by adaptively weighting it with the previous mean. The variance of the filtered signal is computed every 500ms and afterward compared to the variance threshold. Figure \ref{fig:signalProcessing} illustrates the procedure with example values. The signal quality algorithm meets three criteria:

\begin{itemize}
    \item It does not rely on pre-processed information of the recording hardware: The algorithm relies on the variance of the raw signal, and a threshold that is determined a-priori.
    \item It is lenient at the beginning of the procedure to motivate subjects: The algorithm employs a moving average to filter the raw EEG time-series, weighted by the previous signal quality estimate. A drop in signal quality will reduce the influence of the next EEG sample, and priority will be given to the previously computed average. This reduces the negative effects of sudden movements or an overall bad fit of the recording hardware without freezing at 0\% to keep subjects motivated. At higher signal quality, the next EEG sample will be weighted more than the previous average. The variance computation will more accurately reflect the variance of the raw EEG signal. The procedure becomes less lenient to ensure that subjects can only finish fitting if the headset has been carefully adjusted, and the variance of the unfiltered EEG signal stays below the set threshold for several computation cycles. 
    \item The computation is sufficiently fast to be performed on the device alongside any other parallel processes. In the pilot study, the battery life of the iOS device remained sufficient for approximately three hours of recording while running this algorithm in the background and showing real-time feedback.
\end{itemize}

\subsection*{Recording}
After the EEG headset is fitted properly, the recording session starts. Instructions are visually and verbally presented to the subjects in fixed periods with short breaks in between. The EEG time-series is recorded in the background, and phases are marked in the dataset. 

A scenario is split up into smaller blocks of approximately six minutes. After each block, subjects can review their progress, and choose to continue with the next block or end the recording session. 
The fitting screen is presented again before subjects can continue with the next block to ensure good signal quality.
As subjects keep the headset on their head and the quality of all sensors is presented simultaneously, the check-up is typically fast:
Previous results from \citep{Hohmann2019} show that the average check-up was completed in 21 seconds, while the initial fitting required 68 seconds.

\subsubsection*{Questionnaire data}
Questionnaire support was implemented by utilising ResearchKit. 
Questionnaires can include multiple-choice questions, short text answers, and rating scales. 
All items are defined in one JSON file per questionnaire and supported language. 
During the scenario, each question is presented on the screen with an optional TTS until all questions are answered. 
For questionnaire scenarios, hardware preparation and fitting steps are skipped.

\subsection*{Storage, upload and completion}
 If the subject chooses to end recordings, or if all sessions for the day are completed, data is stored and marked for upload. If an internet connection is available, the upload procedure is initiated with a visual representation. The current recording is uploaded together with any previous recording that had not been transmitted. Afterward, the subject is taken back to the home screen.
 
EEG recordings are stored in the widely-used HDF5 standard, and questionnaire results are stored in a JSON file. Both include timestamps, locale, and other meta-information, like fitting time or sensor locations of EEG recordings. As the recorded data is sensitive, we employed asymmetric encryption of all data that is recorded by the application. A public key is stored on the device for encrypting files before upload. Only the experimenter owns the private key to decrypt the data. Additionally, all data is labeled with a unique ID that is randomly generated when the participant signed the consent form. The real name of the participant is only stored on the device and on the consent form. 
The dataset itself is anonymized. 
Storage and transmission are compliant with the General Data Protection Regulation (GDPR).

\begin{figure*}[t]
  \centering
  \includegraphics[width=17.8cm]{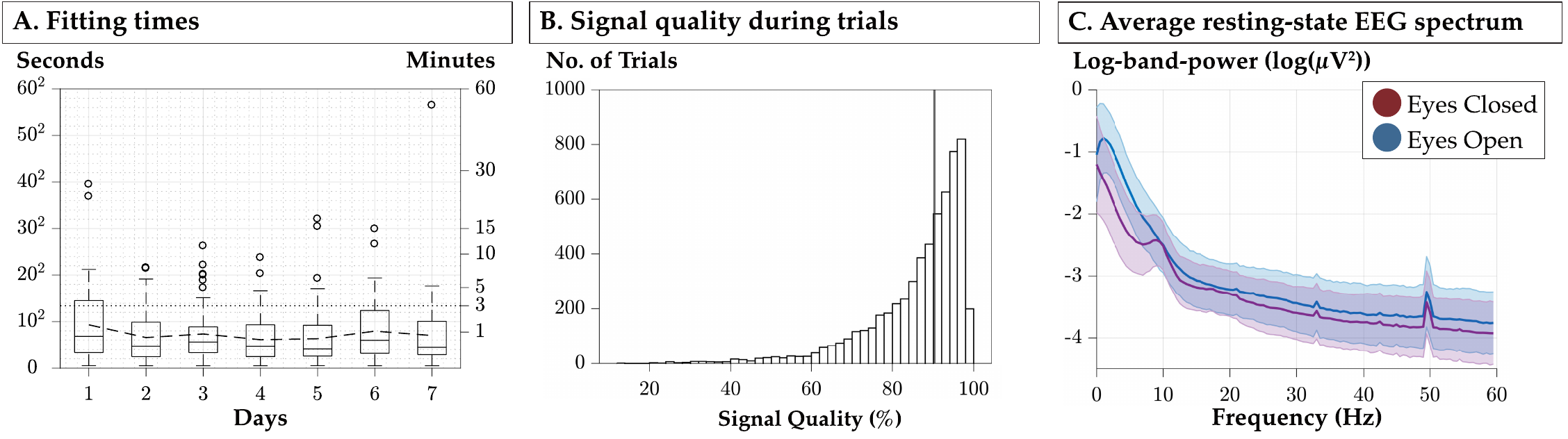}
  \caption{(A) Fitting time of the EEG headset by day. The dotted line represents the three-minute mark, after which the fitting criterion was lowered. Black bars indicate the median, dashed lines indicate the mean. Boxes extend between the 25th and 75th percentile, and whiskers extend to $\pm1.5$ IQR. Circles represent outliers. Data was square-root transformed for this visualization to improve the distribution symmetry. (B) Histogram of the retained average signal quality per trial across all subjects, days, and scenarios. The black vertical line indicates the median. (C) The parietal resting-state EEG spectrum, averaged over all subjects and trials. Solid lines represent the mean; shaded areas represent the average standard deviation.}~\label{fig:fittingTimes}
\end{figure*}

\begin{figure*}
  \centering
  \includegraphics[width=17.8cm]{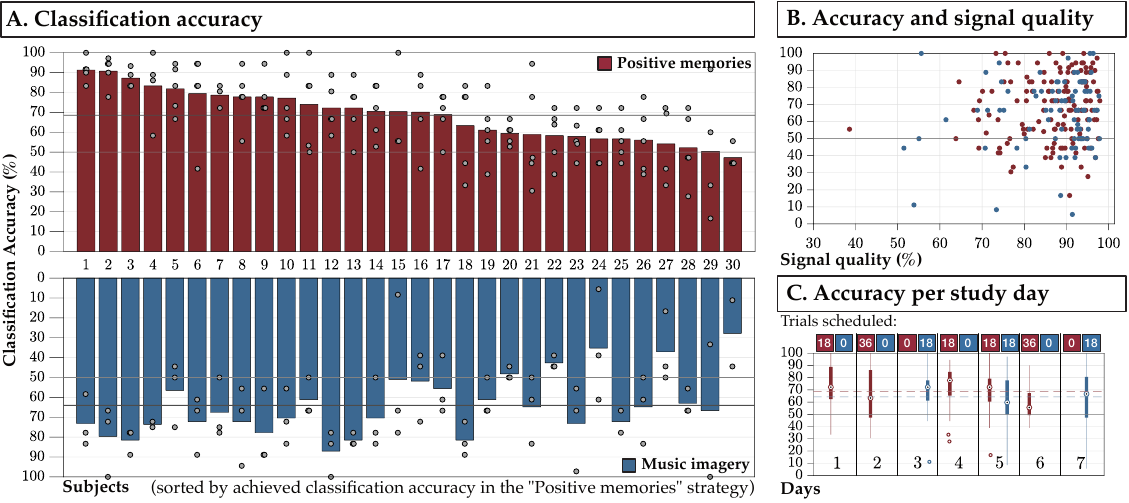} %
  \caption{Results of the at-home evaluation of the BCI control strategies. \enquote{Positive memories} is colored red, \enquote{Music imagery} is colored blue. (A) Mean classification accuracies and signal quality achieved in both strategies. Black lines indicate the mean accuracy. Grey circles indicate accuracies per day.  (B) Daily classification accuracy by achieved signal quality and strategy. (C) Classification accuracy by day and strategy. Dashed colored lines indicate the mean for each strategy. Boxes extend between the 25th and 75th percentile, and whiskers extend to $\pm1.5$ IQR. Circles represent outliers. Gray lines indicate chance-level of 50\%.}~\label{fig:differences}
\end{figure*}

\section*{Results}
Concerning the usability of the system, figure \ref{fig:fittingTimes} (A) shows the average fitting time of the EEG headset across all subjects for each day. The median fitting time across all days and subjects was 25.9 seconds.  
16.2\% of all fitting procedures required more than three minutes, after which the required signal quality dropped to 75\%, as described in the implementation section.
Figure \ref{fig:fittingTimes} (B) shows a histogram of retained signal quality in all trials, with a median signal quality of 90.2\%.
In (C), we show the average resting-state EEG spectrum across all subjects and the average standard deviation.
The dominant frequency between 8 and 13 Hz is visible during the eyes-closed condition.
The spectrum also shows a peak at 50 Hz, likely caused by electromagnetic influence.

Regarding the overall satisfaction with the platform, 25 subjects participated in a post-experiment online survey. 
On the Technology Acceptance Model dimensions \cite[TAM]{Venkatesh2000}, subjects rated perceived ease-of-use with a median score of 5.8/7, perceived enjoyment with a score of 5.6/7, and perceived control over the application with a 6.3/7.
On the NASA Task Load Index dimensions \citep[TLX]{Hart1988}, ratings for frustration, temporal demand, and required effort were low with a median score of 5/21, 8/21, and 7/21.
Subjects rated their median overall satisfaction with an 8/10.

Figure \ref{fig:differences} shows the results of the unsupervised, daily use of both control strategies on consumer-grade hardware.
In (A), we show the average classification accuracies for both strategies in the at-home study, with an overall average accuracy of 68.5\% for \enquote{Positive memories} and 64.0\% for \enquote{Music imagery.}
Grey circles indicate the achieved accuracies per day, where we observe within-subject differences with an average standard deviation of 20.3\%.
We find a small negative trend between mean accuracy and standard deviation (Pearson's $r(58) = -.25, p = .0505$). 
In (B), we relate these accuracies to the signal quality during trials.
We find a small positive trend between daily signal quality and daily accuracies ($r(224) = .13, p = .06$).
We relate accuracies to the day of study in (C).
In general, we find a small negative correlation between the day of study and performance ($r(224) = -.18, p = .006$). 
\enquote{Positive memories} shows a median accuracy of 72.2\% on day 1, and then exhibits a drop in median accuracy to 63.3\% during repeated execution from day 1 to day 2.
After not executing the strategy on day 3, we observe an increase to 77.8\% on day 4, and a subsequent drop to 72.2\% and 55.6\% on days 5 and 6, respectively.
Thirty-six \enquote{Positive memories} trials were scheduled on days 2 and 6, which is the highest daily amount in this study.
In \enquote{Music imagery}, we find a median accuracy of 72.2\% on day 3, where it was introduced as the only scheduled strategy.
Then, we find a decrease to 59.7\% on day 5, where both strategies were scheduled, and a subsequent increase to 66.6\% on day 7, where only \enquote{Music imagery} was scheduled.
With respect to the two potential mediators meditation experience and daily self-reported motivation, we found that meditation experience is unrelated to accuracies ($M = 1.80/3, SD = 0.56, r(224) = .03, p = .65$). 
Motivation correlates positively with accuracy in this study ($M = 3.98/5, SD = 0.92, r(224) = .18, p = .006$).
Subjects recorded 70.4 hours of EEG data. 
Median trial completion was 91.0\% $\pm$ 9.0 MAD, and the median time window between the first and the last recording was 7.8 days $\pm$ 1.7 MAD.

\section*{Discussion}
We developed the \textit{MYND} framework to evaluate laboratory-based BCI control strategies in an unsupervised, realistic scenario. 
Thirty subjects used the application over seven days at home and executed the two control strategies \enquote{Positive memories} and \enquote{Music imagery.}
In terms of platform usability, subjects retained a high signal quality after self-administered fitting with a median of 90.2\%. 
They required below one minute on all days for average preparation time.
Post-experiment survey results indicate that subjects were overall satisfied with the application.
Concerning the two BCI control strategies \enquote{Positive memories} and \enquote{Music imagery,} induced differences in neural activity could be decoded with an average accuracy of 68.5\% and 64.0\% across days, respectively.
Our results indicate that combining a consumer-grade EEG with guided self-administration through \textit{MYND} could be a promising basis to evaluate laboratory-based BCI control strategies for unsupervised, daily use.
Apart from the general evaluation of both strategies in this context, \textit{MYND} allowed us to investigate how signal quality and the repeated execution of both strategies across days may have mediated performance.

Interestingly, maintaining a high signal quality, i.e., a low signal variance, after self-supervised fitting did not necessarily lead to high task performance in this study.
On the other hand, we found a high variation of daily accuracies within subjects.
The consumer-grade EEG and our fitting algorithm allowed subjects to record neural data quickly, but the hardware may also pose new challenges for the BCI user: 
Even when perfectly fitted, the dry sensors at distant ends of the scalp may require users to induce strong, consistent modulations in neural activity in order to be reliably detected.
As one potential mediator, meditation experience may affect the ability to maintain focus in unsupervised environments \citep{Lutz2008}. 
However, in a post-hoc analysis, we found that self-reported meditation experience is unrelated to accuracies in this study. 
On the other hand, daily motivational scores correlate positively with accuracy in this study.
Therefore, using the \textit{MYND} platform to combine easily accessible control strategies with motivating, daily feedback could further improve performance and help with more consistent execution.
This feedback will need to be robust to the lower signal-to-noise ratio and higher susceptibility to involuntary head-muscle artifacts around the ears and the forehead, where the sensors are located. 
As one promising approach to implement robust feedback, concurrent work in \citep{Kobler2019} explored an adaptive signal-filtering method that relies on signal-variation, similar to the fitting algorithm presented here.
Providing immediately accessible daily decoding scores could be implemented on the device by utilizing prior information from higher quality laboratory data via transfer-learning, as shown in this offline analysis.

Both control strategies were immediately executable on the day of introduction, with a median decoding accuracy of 72.2\% each. 
While an accuracy of $\geq70\%$ is typically considered sufficient for BCI control, the study with \textit{MYND} also revealed potential challenges for the reliable, every-day use of the two strategies in future BCIs:
Apart from within-subject variation, our at-home results indicate that repeated task execution across days may have had a negative effect on decoding accuracy. 
Conversely, we found an increase in accuracy after not executing one strategy on day 3, and the effectiveness of the two strategies differs in several cases, for example, in subjects 18, 23, and 25.
Longer, more complex study protocols may be needed to detail the interactions between BCI control strategies, subjects, and time.
However, these first observations may already have implications for future work with \textit{MYND} on enabling reliable, long-term BCI control:
Self-devised strategy schedules and daily feedback could enable subjects to switch between control strategies and devise new strategies that work particularly well for them.
This may reduce the effects of repeated execution of a single strategy and potentially reduce within-subject variation as well.
Subjects could indicate if their strategy choice was influenced by external factors, if they switch strategies after repeated use, or if they stop using concrete strategies altogether after repeated training.
Participants could take on a more active role when investigating the efficacy of BCI control strategies by reporting personal strategies, use means of continuous self-evaluation, and share experiences with other subjects, all of which may positively affect performance by adding potential sources of intrinsic motivation \cite{Morschheuser2017}.
Using consumer-grade hardware and the \textit{MYND} application, adding a set of strategies as a starting point for everyone, and using feedback and personal schedules to improve the ability to elicit neural modulations may be a viable basis for research on accessible, reliable, long-term BCI usage in daily life.

\matmethods{

We first motivate the choice of the two BCI control strategies and the recording hardware.
Then, we outline the experimental procedure.
Afterward, we present the analysis of the usability of the \textit{MYND} platform and the BCI control strategy evaluation.

\subsection*{Choice of laboratory-based BCI control strategies}
We utilized a set of stimulus-free two-task BCI control strategies that were previously evaluated in laboratory experiments with healthy subjects.
The control strategies have shown to be immediately executable by subjects without the need for prior training, and they are unrelated to motor processes, which limits accidental movement artifacts. 
Most importantly, they induce modulation of neural activity in broad areas of the cortex, which could make them particularly suitable for use with consumer-grade EEG equipment with few sensors.
We could show that both BCI control strategies could be used in a laboratory study to induce significant differences in neural activity \citep{Hohmann:vc, Hohmann2016, Jayaram2017, GroHohPetGro17}.  
Now, we complement the laboratory results with \textit{MYND} to investigate whether these strategies can be executed in an uncontrolled, realistic environment, by utilizing a consumer-grade EEG, and if repeated execution affects performance.

In the \enquote{Positive memories} strategy, subjects are asked to switch between thinking about a positive memory of their past or consecutively subtract a small number from a larger number \citep{Hohmann:vc, Hohmann2016, Jayaram2017}. 
Changing between daydreaming and a task that requires attention modulates activity in the \enquote{default mode network,} a large-scale network that is involved in self-referential processes \citep{Fomina2015, Raichle2001}. 
It was found that thinking about positive memories increases activity in the alpha-band, while mental calculations decrease activity. 
Similarly, the \enquote{Music imagery} strategy instructs subjects to switch between playing their favorite song in their head or consecutively subtract a small number from a larger number, which has also been found to modulate parietal alpha-activity \citep{Schaefer2011, GroHohPetGro17}. 
Self-referential thoughts have shown to modulate activity in the theta- (3--7~Hz) alpha- (8--13~Hz) and beta-bands (17--30~Hz) of the human EEG \citep{Fingelkurts2012, Mu2010}.
Further, it was found that the dominant frequency of the EEG spectrum, the \enquote{alpha peak frequency,} is modulated by these strategies as well \citep{Jayaram2017}.
In figure \ref{fig:topoplots}, we used the laboratory data from previous studies \citep{Hohmann:vc,GroHohPetGro17} to illustrate the induced band-power modulations parietal and prefrontal cortex with the two BCI control strategies. 
We computed the coefficient of determination ($R^{2}$) per subject and channel to visualize the average induced differences in normalized theta- alpha- and beta-bands, as well as the dominant frequency of the EEG spectrum.
Lab recordings used a 128-channel \enquote{BrainAmp} EEG system (Brain Products GmbH, Germany) with wet electrodes in a single-day recording. 
Eleven participants completed 40 trials of \enquote{Positive memories} \citep{Hohmann:vc}, and 10 different participants completed 20 trials of \enquote{Music imagery} \citep{GroHohPetGro17}.

\begin{figure}[h]
  \centering
  \includegraphics[width=\columnwidth]{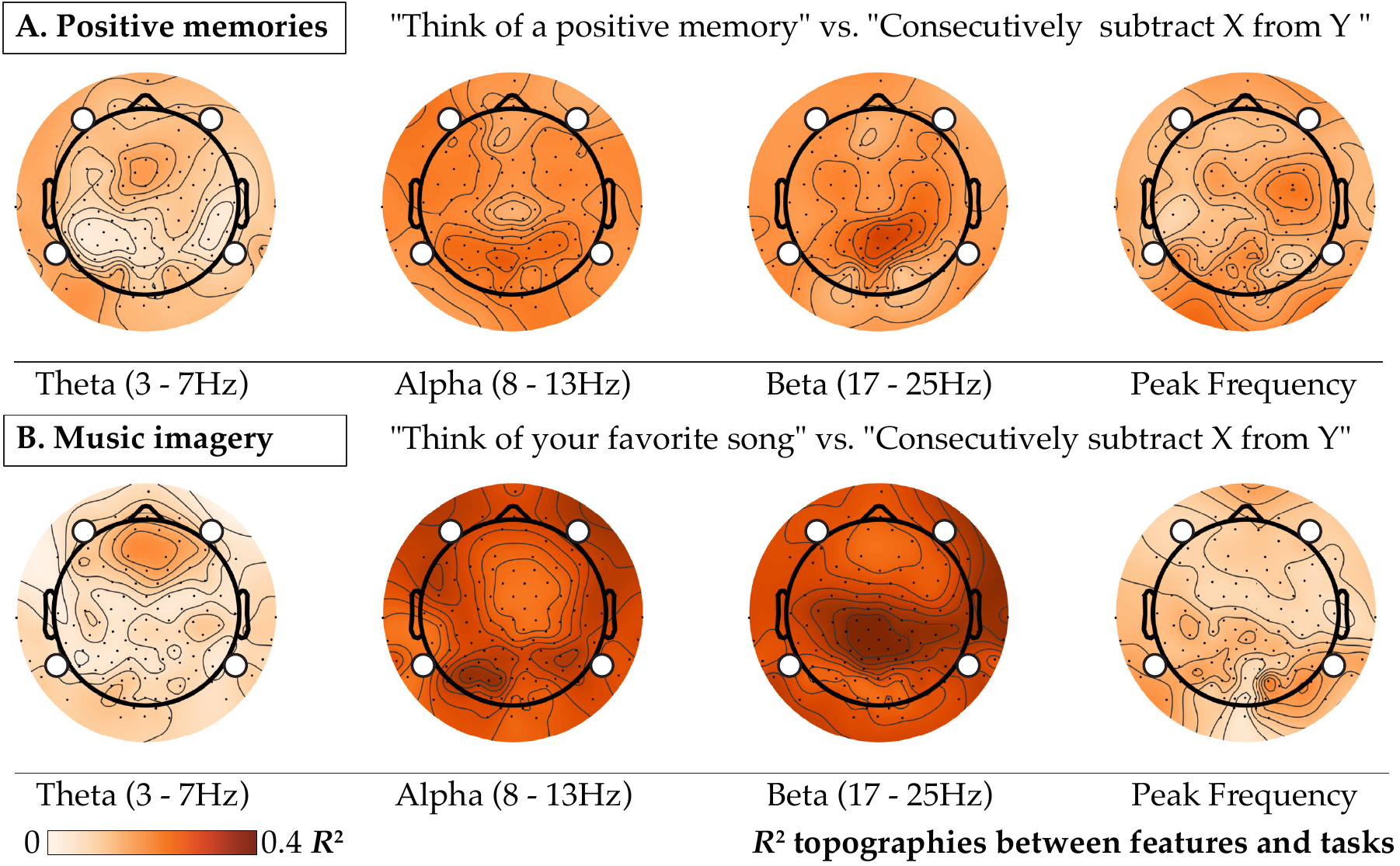}
  \captionof{figure}{Average cortical $R^{2}$-maps of the BCI control strategies \enquote{Positive memories} (A) and \enquote{Music imagery} (B), based on laboratory data from \citep{Hohmann2016} and \citep{GroHohPetGro17}.  White circles represent the sensor locations of the employed consumer-grade EEG headset in the current study.}
  \label{fig:topoplots}    
\end{figure}

\subsection*{Choice of consumer-grade recording hardware}

We chose the Muse EEG headset (2016) by InteraXon as recording hardware.  
The Muse features a rigid plastic headband design and four dry EEG sensors at AF7, AF8, TP9 and TP10 of the International 10--20 system \citep{jasper1958ten}, and a 256~Hz sampling rate.
Figure \ref{fig:topoplots} illustrates the sensor locations.
Despite possible improvements in future sensor technology, the headband-like shape of the Muse may approximate the overall design future consumer-grade EEGs well with respect to a potential integration into glasses or headphones, instead of extensive sensor placements across the cortex.
The Muse EEG was designed to be worn for everyday meditation assistance, and the software library allows for communication with the headset via Bluetooth LE to access to the raw data in real-time. The Muse was previously used in a laboratory BCI speller system in \citep{Krigolson2017}.

\subsection*{Apparatus and participants}
We conducted an evaluation study with 32 subjects who used \textit{MYND} over seven days at home. 
Subjects were recruited via social media from both local and remote communities within Germany. 
Local subjects were able to pick up the equipment at the institute, while a return-shipment of the equipment was provided for remote subjects.
Of the 32 participants, two subjects dropped out of the study before reaching the final day of recordings due to time constraints and the inability to fit the headset to their head, respectively.
This left 30 subjects for the analysis (age 32.5 $\pm$ 9.0 years, 19 female).
Nineteen subjects chose the English version of the application. 
The \textit{MYND} iOS app was installed on 30 iPad (2018) devices, running iOS 12.1.

\subsection*{Experimental procedure}
A package consisting of an iPad with \textit{MYND} installed, a Muse EEG headset, respective chargers, and a small printed manual explaining general usage of the iPad were handed out or shipped to 32 subjects.
Subjects were free to start the seven-day study at any time by completing the boarding procedure. 
Afterward, subjects were asked to complete the respective scenarios for every day, constrained by the time-limit as described in the implementation section.
After the last day, subjects were asked to return the equipment in person or via postal service and fill out an online survey about their experience. 
As in the previous laboratory study, subjects were reimbursed with 12 Euro per hour of system usage, and no additional feedback on performance was given.

Subjects were asked to complete 42 60-second trials of the \enquote{Resting-state} strategy, 126 30-second trials of the \enquote{Positive memories} strategy (on days 1,2,4,5, and 6), and 54 30-second trials of the \enquote{Music imagery} strategy (on days 3, 5, and 7) during the course of the at-home study. 
All strategies were split into blocks: one trial per task for \enquote{Resting-state,} and three randomized trials per task for \enquote{Positive memories} and \enquote{Music imagery.} 
We altered control strategies and the number of trials that subjects were asked to perform on each day to vary between days of lower and higher workload. 
The schedule also allowed us to observe potential effects of repeated usage, breaks, or simultaneous scheduling of both strategies.
Subjects were asked to perform all tasks with closed eyes to prevent eye-blink artifacts.
At the beginning of each day, subjects would rate their motivation to record the given sessions on a 5-point Likert scale (\enquote{How motivated are you to complete today's sessions?}). 
Meditation experience was recorded on the first day of the study on a 3-point scale. 

\subsection*{Analysis} 
All analyses were performed offline in Matlab 2019b (The MathWorks Inc., USA). 

\subsubsection*{Platform usability}
Concerning the usability of the system, we first analyzed the time that subjects required to fit the headset at the beginning of every recording session. 
We reconstructed the retained signal quality during trial execution with the same processing pipeline as described earlier, and we plotted the \enquote{Resting-state} EEG spectrum, averaged over both parietal channels, for a visual inspection of the recorded data. 
During the \enquote{Resting-state,} subjects are asked to either open or close their eyes and let their mind wander. 
Alpha-band-power increases visibly during closed-eyes resting. Therefore this strategy is often used as a benchmark for EEG recording setups (e.g., \cite{Grosse-Wentrup2014a}).
Concerning the subject's satisfaction with the system, we present the post-experiment ratings of the application in an online survey, to which 25 subjects responded.
We report the dimensions \enquote{perceived ease-of-use}, \enquote{perceived enjoyment}, and \enquote{perceived external control} of the Technology Acceptance Model \citep[TAM]{Venkatesh2000}, as well as \enquote{temporal demand}, \enquote{frustration}, and \enquote{effort} of the NASA Task Load Index \citep[TLX]{Hart1988}, and a 10-point total satisfaction score, as proposed in \citep{Kubler2014} for BCI development.

\subsubsection*{Control strategy evaluation}
With this study, we aim to evaluate how well the BCI control strategies can be performed in an uncontrolled, realistic setting and consumer-grade hardware, outside of the laboratory.   
The small amount of trials scheduled per day (18 to 36), recorded with a consumer-grade EEG, makes it challenging to learn a good decoding model per subject, day, and strategy.
To solve this problem, we leveraged the information obtained from previous laboratory trials with a transfer-learning approach (\cite{Jayaram2015}, toolbox available online\footnote{https://github.com/vinay-jayaram/MTlearning}, additional details in \cite{Hohmann2016}).
Here, a linear regression model for each subject in the at-home study is learned while regularizing the regression weights with a Gaussian prior that is learned on existing laboratory data.
The three-step implementation, which we describe below, also resembles a possible procedure for daily on-device feedback that would be immediately accessible to subjects in future studies.

First, we performed the following EEG preprocessing steps per subject, day, and strategy, to obtain the feature space for pattern classification:
We windowed the EEG time-series at every channel with a Hann window and computed the log-band-power for every trial for both laboratory studies and the at-home study.
Muscular artifacts in the laboratory studies were removed from the data with an independent component analysis (ICA) before band-power computations, as described in \citep{Hohmann2016}.
We extracted four features: theta- (3--7~Hz) alpha- (8--13~Hz) and beta-log-band-power (17--30~Hz), as well as the dominant frequency of the EEG spectrum as described in \citep{Jayaram2017}.
Then, we used mean and standard-deviation to normalize band-powers and the dominant frequency for every subject across the whole session in the laboratory study, and within each day in the at-home study. 
We extracted theta-, alpha-, and beta-band-power, and the dominant frequency at the electrode locations of the consumer-grade headset (TP8, TP10, AF1, and AF2, see figure \ref{fig:topoplots} for illustration), resulting in a total of sixteen features per trial and control strategy.

Second, we learned a Gaussian prior over all subjects of the laboratory dataset.
This laboratory prior is updated in two steps:
First, its covariance is kept constant, and regression weights are computed for a given subject.
Second, the covariance of the prior is updated with the weights of the learned model.
We used the \enquote{convex multi-task feature learning} algorithm as update method \citep{Argyriou2008}, and iterated the procedure 10,000 times.

Third, we learned decoding models per subject, day, and control strategy in the at-home study in a leave-one-trial-out cross-validation procedure.
For $n$ trials per subject, day, and control strategy, we fit a linear regression model on $n-1$ trials, regularized by the previously obtained prior.
Then, we used the learned model to predict the task label of the $n$-th trial, resulting in one classification accuracy per subject, day, and strategy.

We related the obtained accuracies to four potential mediators of decoding performance: Average signal quality during trials, the day of the study, meditation experience, indicated at the beginning of the study, and daily self-reported motivation. 
}

\showmatmethods{}

\acknow{We would like to thank Florian Remele, Timothy Gebhard, Maria Wirzberger, Ann-Kathrin Zaiser, and Vinay Jayaram for their help with preparing this manuscript, and Hubert Jacob Banville for their suggestions on headset connectivity improvements.}

\showacknow{} %
\printbibliography

\end{document}